\documentclass[aps,prl,twocolumn,showpacs,preprintnumbers]{revtex4-1}

\usepackage{psfrag,graphicx}
\usepackage{dcolumn}
\usepackage{amsmath,amssymb}
\usepackage{bm}
\usepackage{amsfonts,amssymb,amsmath}        
\usepackage{epstopdf}
\usepackage{natbib}
\usepackage{amsthm}
\usepackage{hyperref}
\usepackage{xcolor}

\newcommand{\be}{\begin{equation}}
\newcommand{\ee}{\end{equation}}
\newcommand{\bq}{\begin{eqnarray}}
\newcommand{\eq}{\end{eqnarray}}

\newcommand{\ket}[1]{\left | \, #1 \right\rangle}

\begin{document}

\title{Teaching quantum computing with an interactive textbook}
\author{James R. Wootton, Francis Harkins,  Nicholas T. Bronn, Almudena Carrera Vazquez, Anna Phan, Abraham T. Asfaw}
\affiliation{IBM Quantum}
\date{\today}

\begin{abstract}
Quantum computing is a technology that promises to offer significant advantages during the coming decades. Though the technology is still in a prototype stage, the last few years have seen many of these prototype devices become accessible to the public. This has been accompanied by the open-source development of the software required to use and test quantum hardware in increasingly sophisticated ways. Such tools provide new education opportunities, not just for quantum computing specifically, but also more broadly for quantum information science and even quantum physics as a whole. In this paper we present a case study of one education resource which aims to take advantage of the opportunities: the open-source online textbook `Learn Quantum Computation using Qiskit'. An overview of the topics covered is given, as well as an explanation of the approach taken for each.
\end{abstract}

\maketitle

\section{Introduction}

Quantum information is an interdisciplinary field, primarily associated with the traditional disciplines of physics, mathematics and computer science. The teaching of quantum information science is strongly influenced by this, with introductory courses typically assuming an audience of either physicists, mathematicians or computer scientists, and tailoring the content accordingly. For the physicists, there has also traditionally been a large difference between theoretical and experimental approaches: the former focussing on the theory and applications, whereas the latter focussing on what it means to build a quantum bit.

The many advances made in quantum computation in recent years motivate the development of a new approach to education in this field. One that does not assume a specific background, and which presents topics more in terms of software and hardware than theory and experiment. The open-source online textbook `Learn Quantum Computation using Qiskit'\cite{qiskit-textbook} is one example of the approaches taken over the last few years~\cite{Svore2020,Economou2020} to address this issue.

The textbook is inextricably linked with quantum hardware and software, allowing a hands-on approach. It is tailored to the hardware provided by the \emph{IBM Quantum Experience}~\cite{iqx} which, at time of writing, provided public access to devices of up to 15 qubits via a cloud service. The software framework for quantum computation that it is tailored to is, of course, \emph{Qiskit}~\cite{qiskit}. This is an open-source Python-based framework maintained by IBM Quantum.

The initial construction of the textbook drew inspiration from two main sources. One is the set of tutorials provided as part of the IBM Quantum Experience at its launch in 2016~\cite{iqx-guides}. The other was a course given by one of the authors at the University of Basel. This course was offered by the physics department for undergraduate and masters students, but could also be attended by students from other courses, including nanoscience and computer science. It was therefore conceived as an interdisciplinary course, taking students from both a physics and computer science background and teaching them a topic which bridges the two. The course was developed over several years before the textbook was written, had many important elements written while teaching that course during 2019, and was used as the basis for the course in 2020.

Given this origin, the textbook has been designed both to enable self-study, and to provide a resource to educators providing a university-level course on quantum computation. When discussing the intended audience of the textbook in this paper, we will typically refer to a self-studying `reader'. However, the same considerations will also apply to a student using the textbook as part of a course.

In this paper we summarize the content and approach taken in each part of the textbook, typically with one section devoted to each chapter.

\section{Chapter 0: Prerequisites and Chapter 8: Appendix}

The only prerequisite required to follow the book from start to finish is a basic familiarity with Python and Jupyter notebooks. For readers who do not already have this knowledge, a summary of the important concepts is provided in the initial Chapter 0.

For readers who prefer to skip to later chapters, such as those on quantum algorithms, an appropriate knowledge of linear algebra is very much a requirement. A complete account of all the linear algebra concepts that may be required is therefore provided in the appendix. This is available for the reader to consult before dipping into later chapters, or to refer to as and when necessary.

\section{The opening sections}

Introductions to quantum computing often begin with an equation such as
\be
a \ket{0} + b \ket{1}
\ee
This short equation contains a lot to unpack. The $a$ and $b$ are complex numbers. The $+$ refers not to the familiar addition of primary school, but to addition of vectors within linear algebra. The $\ket{0}$ and $\ket{1}$ are unit vectors which, even for a reader who is familiar with such concepts, are expressed using a very niche notation.

Since this equation is so dense with technicalities, and so full of assumptions regarding prior knowledge, it was decided to use it sparingly within the first sections of the Qiskit textbook. Instead, the aim is to begin in a way that is more suitable to the many people who have an interest in learning more about quantum computing, but whose knowledge of complex numbers and linear algebra is non-existent or long-forgotten. Though this is not the main target audience of the textbook as a whole, it was nevertheless regarded that all audiences might benefit from the establishment of intuition and motivation before moving on to technicalities.

With this in mind, the aim of the beginning of the textbook is to do the following:

\begin{itemize}
\item Establish how quantum computation differs from standard digital computation.
\item Give the reader enough information to do something interesting with Qiskit, or the GUI interface of the IBM Quantum Experience.
\item Do the above without using or assuming knowledge of linear algebra or complex numbers.
\end{itemize}

These aims are specifically addressed in the two opening sections of the textbook, \emph{The Atoms of Computation} and \emph{What is Quantum?} .

\subsection{The Atoms of Computation}

To achieve the first aim there is an important issue to tackle: most people do not have a concrete idea of what conventional digital computation is. They may have a concept that it takes in information, processes it and then outputs information. They may also know that the information is encoded in binary. However, this simple
description arguably holds just as true for many quantum algorithms as it does for convention digital computation. The difference is in the details of the process, and the basic operations into which all tasks can be compiled. Most audiences will not be used to thinking of computation at this `machine level' description, in terms of Boolean logic gates. Whether the reader is an undergraduate physicist, computer scientist or a motivated layperson, it is necessary to get them into this mindset before beginning.

A consequence of the third aim, particularly the avoidance linear algebra, is that the notion of \emph{superposition} should be avoided at this initial stage. Though often used in popular science accounts in hand-waving explanations of quantum states being `two things at once', superposition specifically refers to the linear combination of state vectors (or wave functions). Without linear algebra, and without wanting to adopt a hand-waving approach, this term must be avoided until the relevant mathematical tools are introduced.

With these points in mind, the first section of the Qiskit textbook is `The Atoms of Computation'. This seeks to get the reader thinking in terms of algorithms using the example of the standard addition algorithm. Even a reader who thinks they know nothing of algorithms should certainly recall this process of adding single digit numbers in order to add numbers of arbitrary size. This familiar and accessible algorithm is then cast into binary in order to introduce the most important Boolean logic gates from the perspective of quantum computation: the \texttt{XOR} and the \texttt{AND}. These logic gates are expressed using their quantum equivalents (the controlled-NOT and Toffoli, respectively) as part of quantum circuits. At the end of this section, the reader should then have the ability to understand the basics of classical computation, of binary and of expressing algorithms through gates, and even be able to reproduce simple classical logic gates using Qiskit.


This section ends with a short justification of how quantum computation will differ from these simple Boolean logic gates. However, it is not sufficient to fully satisfy our first aim. To do this we must first ask the reader to choose how to proceed: with or without linear algebra. Those who choose linear algebra should simply proceed on to the next section. By the end of the first chapter, they should have a firm grasp of the basics of quantum computation, and how it differs from classical computation. For those who are not yet ready for linear algebra, we must find an alternative means to represent simple quantum states and gates. This is done in \emph{Hello Qiskit}, which provides an interactive, game-based way for readers to become familiar with the basic principles of quantum computing. Both these approaches are expanded upon in their corresponding sections below.


\subsection{What is Quantum?}

\emph{What is Quantum?} is a short, standalone chapter aimed at beginners who are unfamiliar with quantum computing and are considering learning more. Unlike other texts aimed at similar audiences, this chapter uses mathematics to describe the behaviours behind quantum computing. By including a mathematical approach, which is nevertheless constrained to concepts that should be familiar to all readers, this chapter aims to provide readers with a concrete picture of the quantum phenomena exploited in quantum computing, and an idea of what the rest of the textbook involves.

The chapter builds upon Scott Aaronson's approach to introducing quantum mechanics, with the ``conceptual core'' of quantum mechanics being ``a generalization of probability theory to allow minus signs''~\cite{aaronson}. Aaronson's text requires the reader have a certain level of mathematical maturity (e.g. the reader should be comfortable with complex numbers and vectors), but \emph{What is Quantum?} requires only familiarity with probability trees, multiplication and square roots.

Additionally, the chapter aims to familiarise the reader with the scientific process of using a model to describe behaviours and modifying that model when it is unsuccessful.

The chapter starts by reminding the reader of probability trees. The chapter shows the reader how we can create a simple mathematical model of a coin toss using these trees and how we can use this model to predict the results of different experiments. The interactive nature of the textbook allows the reader to simulate these experiments in their browser and confirm these predictions.

The chapter then introduces the qubit as the `quantum coin', i.e. a coin that follows the rules of quantum mechanics. The Hadamard gate (referred to as the `quantum coin toss') is introduced, and the reader encouraged to experiment with the quantum coin and use probability trees to build a model of its behaviour. Upon the application of two Hadamard gates, quantum interference is observed, and probability trees can no longer describe the behaviour of the quantum coin. The reader is introduced to quantum mechanics as ``probability theory with negative numbers'' which then allows us to build a working model of the quantum coin.


\section{Chapters 1 and 2: Introducing required tools and techniques}

The aim of the first two chapters is to introduce everything that a reader will need to know in order to understand quantum algorithms. These are the core of the textbook, and the parts that were most heavily based on the original IBM Quantum Experience and the University of Basel course.

\subsection{Chapter 1}

Following on from \emph{The Atoms of Computation}, the remaining sections of Chapter 1 are written with the assumption that the reader has been taught linear algebra, but has not used those skills in some time. This will act as a compromise between the various different audiences, in order to not be too slow for those who are well-practiced experts at linear algebra, and yet still be accessible to a sufficiently motivated newcomer who aims to the learn the required maths along the way. This approach is also taken to account for the fact that quantum computing uses its own unique dialect of linear algebra, with its own notation and a particular focus on certain methods and techniques. It is possible for a reader's linear algebra courses not to have prepared them for the particular ideas that are used. This is something that has been seen in many years of teaching interdisciplinary courses on quantum computing at the University of Basel: almost all students had taken linear algebra courses, but that did not mean that every use of linear algebra in the course was second nature to them. Introducing the mathematical concepts as we go along therefore provides the most accessible introduction.

In order to accommodate those readers who do not have complete comfort with linear algebra, the first two chapters of the textbook teach readers just enough linear algebra needed to cover the content of the algorithms in Chapter 3. In teaching just enough, vectors are introduced only as lists of complex numbers, and more generalised ideas about vectors are ignored. It is hoped that the behaviour of vectors and matrices as seen through the behaviour of qubits and gates will help students when learning about more general vector spaces.

To clarify the separation between mathematical prerequisites and concepts that should be new to the reader when learning about quantum computing, the linear algebra material has been enclosed in expandable accordions, with the content hidden by default. Expandable accordions allow the readers that are comfortable with linear algebra to avoid recapping familiar material and avoid convoluting new content in the chapter.

After two sections which introduce how quantum states and gates can be expressed mathematically, Chapter 1 ends by introducing another important mathematical tool: complexity theory. This is the language in which classical and quantum computation are typically compared, and in which the notion of a quantum speedup is formulated. To introduce the required concepts from this field in an approachable manner, the algorithm for addition introduced in the first section is analyzed in terms of the computational resources it needs.

The chapter ends with a short justification of why quantum computers might offer a speedup over their classical counterparts. This is done in reference to wave particle duality, since this is one of the most well-known parts of quantum mechanics (even to a beginner). With this we can express that quantum computation is neither of the standard forms of classical computation: digital or analog. Rather, the idea of wave-particle duality translates into a form of digital-analogue duality. Though admittedly this is a relatively hand-waving explanation, it is beyond the scope of the first chapter to fully explain the workings and potential of quantum computers. Instead, the aim of this explanation is to disabuse the reader of mistaken notions that they may have. Education in any part of quantum mechanics brings many misconceptions~\cite{Styler1996,Passante2015,Modir2019}. Typical misconceptions specific to quantum computing are that a quantum advantage is possible somehow because of the randomness of quantum mechanics, or because superposition allows exponential parallelism, or that it is a form of computation that is entirely different from those we already know. Instead it shows that quantum computation is not simply a quantum enhancement of classical methods, but that it is nevertheless a cousin to them.

With this, the reader should be ready to find out exactly what quantum computation is.

\subsection{Chapter 2}

The aim of Chapter 2 is to introduce sufficient tools and ideas to be able to explain and prove universality as it relates to quantum computing. Universality shows that quantum gates can perform any computation, and explicitly defines what it means for a quantum computer to do everything a quantum computer can do. This requires an understanding of multiqubit states and gates, the nature and role of entanglement, as well as the means by which gates can be combined to create complex effects.

Universality is proven by introducing the unitary and Hermitian matrices that are so important in quantum computing. Specifically, unitaries are shown to be the way in which inputs are mapped to outputs, and that performing any possible computation means realizing any possible unitary. The Trotter-Suzuki method is then used to show how the single and two qubit gates introduced so far can create any arbitrary multiqubit unitary.

One consequence of universality is that a quantum computer is able to reproduce any classical computation, and do so with at least the same computational complexity. This fact is a useful one to point out, since it provides a powerful demonstration of the general applicability of quantum computation. It is also crucial for quantum computation itself, where classical subroutines are used in quantum algorithms. This chapter therefore concludes with a section on this topic, both to reinforce the concept of universality and to prepare for the use of oracles in the next chapter.

Once the reader knows about everything a quantum computer can do, they are ready to understand specific instances. The next step is therefore to look at specific quantum algorithms.

\section{Chapter 3: Quantum Protocols and Quantum Algorithms}

The \emph{Quantum Protocols and Quantum Algorithms} chapter aims to offer an alternative to static textbooks. It guides the reader through the theory of each algorithm (or protocol), and then shows the reader how they can implement each of them in Qiskit. These introductory algorithms focus on building circuits at the gate level to avoid hiding anything from the reader. The text occasionally borrows from Qiskit’s circuit library to simplify some steps and avoid convoluting the demonstrations (e.g. for oracles, or for circuits the reader should already be familiar with), but this is used sparingly.

The chapter begins by explaining the concept of quantum circuits: the central element of any quantum software. This is provided both as a recap for readers who have followed all the chapters so far, and as a necessary prerequisite for anyone who has skipped to this chapter. The chapter then moves on to examples of protocols with features that are unique to quantum information, namely superdense coding~\cite{Bennett1992}, quantum teleportation~\cite{Bennett1993} and quantum key distribution~\cite{Bennett1984}.

At this point, we are able to introduce algorithms for which there is a provable quantum speedup. The two most famous examples are arguably the factoring algorithm by Shor~\cite{Shor1994} and the search algorithm by Grover~\cite{Grover1996}, for which many concrete use cases have been proposed. Before these are introduced, however, simpler `proof-of-principle' algorithms are covered: those of Deutsch-Jozsa~\cite{DJ1992}, Bernstein-Vazirani~\cite{BV1997} and Simon~\cite{Simon1994}. Though these lack the obvious practical uses of Shor's and Grover's algorithm, they provide concrete examples of quantum speedups that are simple to explain and to implement in Qiskit.

The chapter then moves on to Shor's and Grover's algorithms. It starts with the quantum Fourier transform and the quantum phase estimation algorithms that are required for factoring, and concludes with the quantum counting algorithm that builds upon both Grover's algorithm and quantum phase estimation. With these examples and their relationships with each other, the reader should see how quantum algorithms are often built from clever combinations of known techniques.

\section{Chapter 4: Quantum Algorithms for Applications}

The \textit{Quantum Algorithms for Applications} chapter is divided into two parts. The first part covers algorithms applied to real world problems, while the second part is focused on more recent developments and is meant to be a living document kept up to date with the field.

Each subsection follows a similar structure, starting by explaining the theory of the algorithm and then showing how to implement it with Qiskit on a specific example, sometimes going as far as running a problem on a quantum device.

The chapter can be viewed as a collection of quantum algorithms which can be read in a nonlinear fashion or skipped altogether. The algorithms presented belong to topics ranging from chemistry to machine learning, and while the textbook tries to incorporate as much background as possible, it cannot be taken as a rigorous introduction to each topic.

The first quantum algorithm introduced is the HHL algorithm \cite{ PhysRevLett.103.150502}, a quantum algorithm to solve systems of linear equations and which is used as a building block of several other algorithms, mostly within machine learning. The main building block of the algorithm is quantum phase estimation, which was already introduced in the previous chapter.

The following subsection explains the Variational Quantum Eigensolver (VQE) \cite{2014Peruzzo} applied to the problem of finding the ground state energy of a molecule. Roughly the idea is to use parametrized shallow circuits and a classical optimizer with a cost function, used to update the parameters of the quantum circuits. The main advantage of VQE is that its shallow circuits can be run on near-term quantum devices, and is therefore a hot area of research.

Afterwards comes the Quantum Approximate Optimization Algorithm (QAOA) \cite{2014Farhi} to solve combinatorial optimization problems, where the goal is to find a solution to a problem that minimizes (or maximizes) some cost function. QAOA can be applied to a wide variety of problems, and in the textbook the reader is introduced to MAXCUT and MAX 3-SAT as illustrative examples. However, it should be noted that this is a heuristic algorithm, meaning that there are no performance guarantees.

In the previous chapter, the reader was introduced to Grover search, and now they are shown to apply it to solve satisfiability problems. Here we are interested on, given a Boolean formula, deciding whether it is possible to replace the variables by the values TRUE or FALSE such that the formula evaluates to TRUE.

The first section concludes with hybrid quantum-classical neural networks, which are implemented by integrating Qiskit and PyTorch, an open-source machine learning framework. The goal of the hybrid approach can be to enhance classical algorithms by outsourcing difficult computations to the quantum computer, or to optimize quantum algorithms using classical machine learning. The purpose of this subsection is to demonstrate the ease of integrating both software packages and encourage research on exploring the possibilities of quantum in machine learning.

Finally, at the time of writing this manuscript, the second part of the chapter contains one algorithm: the Variational Quantum Linear Solver (VQLS) \cite{2019BravoPrieto}. VQLS is a quantum algorithm aimed at solving systems of linear equations, as HHL, but using VQE as its building block. Both HHL and VQLS share the same output, and while HHL obtains a much better speedup than VQLS, the latter can be run on near-term quantum computers.

\section{Chapter 5: Investigating Quantum Hardware Using Quantum Circuits}

Most chapters of the textbook focus on the potential advantages of current hardware. This chapter instead focuses on the imperfections, as well as how to detect them, characterize them and mitigate their effects.

The chapter begins with an introduction to the concept of quantum noise and how to detect and correct it using quantum error correction. This is done using the repetition code, whose classical nature has the advantage of being easy to explain and understand. However, this same nature also has the disadvantage of not being fully useful to detect and correct quantum errors. As such, the repetition code is not introduced as a way to correct errors in itself, but instead as a way to test some of the techniques and assumptions behind all quantum error correcting codes. A variant of this section has been published as a paper in its on right, outlining this approach to benchmarking near-term devices with quantum error correction~\cite{Wootton2020}. Additional sections are planned in future to expand upon quantum error correction and how it enables fault-tolerant quantum computation.

In the near-term, removal of the effects of errors in quantum computation will be done by error mitigation techniques, rather than fully-fledged quantum error correction. The next section is therefore to introduce such a technique as an example. The technique used for measurement error mitigation serves as a useful pedagogical example, since it is essentially based on comparisons of the histograms obtained as the output of quantum circuits. This allows an introduction to the concept of error mitigation, without requiring complex analysis of quantum processes.

The final sections introduce standard techniques for benchmarking quantum hardware: randomized benchmarking~\cite{Magesan2012} and the quantum volume~\cite{Cross2019}. These provide examples of two extremes, with randomized benchmarking providing a characterization of individual gates, and the quantum volume providing an analysis of a device as a whole.

All the topics covered in this chapter correspond to tools available within Qiskit. As well as teaching the topics themselves, these sections also demonstrate how to use the tools in Qiskit. The aim of this chapter is therefore the enable a reader to benchmark quantum hardware for themselves, and to fully understand the results.

\section{Chapter 6: Investigating Quantum Hardware Using Microwave Pulses}

The majority of the textbook explains quantum computation at the gate level. This allows us to focus on the applications to computation, and means we only need to introduce concepts from quantum mechanics that are relevant to the gates being used. The underlying quantum hardware is mostly treated as a `black box', and the physical systems and effects used to create the gates are not explicitly mentioned. This is true even in the previous chapter, where the effects of imperfections in the underlying hardware are discussed and measured within the context of gates.

For much of the target audience of the textbook, this black box approach will be entirely appropriate. However, greater understanding of the hardware will be very important to many readers. The most prominent examples are physicists, whose interest in quantum computing may lie more with the hardware than the applications it can be used for. This chapter is therefore devoted to microwave pulse-level access to quantum hardware, which explains the low level manipulations from which quantum gates are built.

Starting in 2019, Qiskit Pulse provides users of the IBM Quantum Experience microwave pulse-level access to to certain quantum hardware backends~\cite{McKay2018, Alexander2020}. This level of access allows the user to perform the experiments necessary for calibration of transmon qubits~\cite{Koch2007} and explore their device physics in a circuit quantum electrodynamics (cQED) architecture~\cite{Blais2004, Wallraff2004}. This chapter explains these calibration procedures and transmon physics that can be performed with Qiskit Pulse, as well as useful calculations and techniques that are drawn from disparate sources.

The first section describes how to find the transition frequency of a qubit by using a two-tone spectroscopy technique, the physics of which is described later, in which the frequency response of the readout resonator is monitored to determine the frequency at which a qubit pulse excites the transmon from the ground state. This is followed by a Rabi experiment in which the qubit is driven around the Bloch sphere by pulses of different amplitudes to determine that at which the qubit is driven to the $\ket{1}$ state, a so-called $\pi$-pulse because it represents the amplitude at which the qubit is rotated by $\pi$ radians ($180^\circ$) around an axis in the $xy$-plane of the Bloch sphere. The qubit relaxation time ($T_1$) is then determined by an inversion recovery experiment in which the qubit is excited to the $\ket{1}$ state followed by various wait times to determine the residual probability that the qubit remains excited, and $T_1$ is determined as the time constant of this exponential spontaneous decay. Ramsey experiments, which consist of a $\pi/2$-pulse that pulse the qubit into a superposition of $\ket{0}$ and $\ket{1}$ followed by time evolution and another $\pi/2$-pulse that ideally returns the qubit to the ground state, are then described. As Ramsey experiments are sensitive to any change in phase between the $\ket{0}$ and $\ket{1}$ states, this provides information about off-resonant drive (which manifests as an unintentional $Z$-rotation) and dephasing (in the decay constant $T_2^*$). Dynamical decoupling achieved by inserting an echo ($\pi/2$-pulse) halfway during a Ramsey experiment yields the overall decoherence decay constant $T_2$, the standard metric reported by IBM backends. (Note the historical differences between $T_2$ and $T_2^*$ arise from developments in the field of nuclear magnetic resonance~\cite{Vandersypen2005}).

The following section uses \texttt{meas\_level=1} data from Qiskit Pulse to measure and discriminate different transmon states in the $IQ$-plane. This in-phase/quadrature-phase plane is the microwave engineer's version of the complex plane. After determining the qubit frequency and $\pi$-pulse amplitude, as performed in the previous section, a readout discriminator is built from analysis of the measurement points (shots) in the $IQ$-plane by preparing the qubit in the ground or excited state. Similarly, higher levels of the transmon are determined (i.e., the $\ket{2}$ state) by two-tone spectroscopy and Rabi experiments, but with the transmon initially prepared in the $\ket{1}$ state to determine the $\ket{1} \to \ket{2}$ transition frequency and amplitude of the $\pi$-pulse corresponding to that transition.  Analysis of measurements in the $IQ$-plane performed after preparing the transmon in the $\ket{0}, \ket{1}$, or $\ket{2}$ then provides the basis for discrimination between them.

The third section describes the quantization of electrical circuits using the branch-flux method and applies it to both the quantum harmonic oscillator (QHO) and transmon qubit~\cite{Devoret1997, Richer2013, Girvin2014, Vool2017, Krantz2019}. From the quantized Hamiltonians, the difference between the energy levels of the QHO and transmon is made apparent, which provides the transmon the anharmonicity required to address the ground and first excited states as the $\ket{0}/\ket{1}$ quantum computational basis using the \texttt{qutip} package  (which is provided as part of \texttt{pip install qiskit}). Then, neglecting the complications of the cQED architecture for pedagogical reasons, qubit drive and the (commonly-used) rotating-wave approximation are introduced to explain the physics of the Rabi and Ramsey experiments of Section 1. 

The fourth section introduces the Jaynes-Cummings (J-C) Hamiltonian, the simplest description of two interacting quantum systems, in this case a qubit and resonator. This is followed by the Schrieffer-Wolff (S-W) transformation, a common technique to derive an effective Hamiltonian by block-diagonalization~\cite{Winkler2003, Richer2013}. The S-W transformation is then applied to the J-C Hamiltonian to describe the qubit and resonator as separate quantum systems ``dressed" by their interaction. This is presented interactively by using the symbolic Python package \texttt{sympy} (which is also provided as part of \texttt{pip install qiskit}). The S-W is then applied to the full transmon Hamiltonian, but with many of the calculations omitted for brevity, as this goes beyond either the capabilities of \texttt{sympy} and/or the author's ability to use it. A consideration of qubit drive, but now including the cQED architecture neglected in the previous section is then derived. The similar case of applying the S-W transformation to two qubits coupled by a microwave resonator yields the cross resonance Hamiltonian, the native entangling operation for IBM Quantum backends, where it should be noted that full extension to two transmons brings the reader to an area of active research~\cite{Magesan2020, Malekakhlagh2020}. The following two sections explore the effective J-C Hamiltonian derived in the dispersive limit, which explains how the presence of the qubit ``dresses" the resonator and vice-versa. The resonator frequency is measured with a frequency sweep with both a high and low amplitude pulse to determine the linewidth, dispersive shift, and qubit-resonator coupling strength. Conversely, the ac Stark (frequency) shift of the qubit and qubit linewidth broadening is observed as a function of average number of photons in the resonator. 

The final section explores the measurement of interaction rate coefficients in the cross resonance drive via Hamiltonian tomography~\cite{Sheldon2016}.  From the block-diagonal cross resonance Hamiltonian, expressions for the time evolution of operators is derived to explore the Pauli expectation values as a function of time depending on the state of the control qubit, which shows how the target qubit evolves around the Bloch sphere with the application of the cross resonance drive. The Hamiltonian tomography experiment is run on the  \texttt{PulseSimulator} with a Duffing oscillator model built from parameters of a real backend because multi-qubit backends with Pulse access are not openly available at the time of this writing. The interaction rates are then calculated from fits to the simulated data, followed by similar simulation but with a phase shift in the cross resonance pulse to maximize the $ZX$ interaction coefficient. The $ZI$ interaction rate is then determined from a Ramsey experiment to see the ac Stark shift of the control qubit (now due to the cross resonance drive), and is then corrected by the subsequent application of a frame change~\cite{McKay2017}.

\section{Chapter 7: Exercises}

It is usual for textbooks to include exercises which help the reader to test their newly acquired knowledge, and which can be used as the basis for exercise sessions in courses based on the textbook. Since a course on quantum computation at the University of Basel was one of the starting points of the textbook, the exercises from this course were considered as an initial set of exercises for the textbook. However, the majority of these exercises were primarily applications of linear algebra, with no need of or reference to Qiskit. They also could not easily be adapted to an interactive form, which would allow for the textbook to provide feedback on solutions. It was therefore decided that the Qiskit-based and interactivity-compatible exercises would be prioritised. These are the three exercises included in Chapter 7, which are based on the topics of Chapter 1. Two of these use ideas from Section 1.2, challenging the reader to construct classical logic gates from their quantum equivalents. The other is based on ideas of gate synthesis, which are built upon in Chapter 2.

These three exercises are not enough for an entire textbook, or even for Chapter 1 on its own. As such, small exercises are included throughout the main text, as well as an extensive set of puzzles in the \emph{Hello Qiskit} game of Section 9.1. Interactive code examples also allow for experimentation to test out new knowledge. Nevertheless, developing a more thorough approach to exercises is one of the main tasks that will be pursued in the near future.

\section{Chapter 9: Games and Demos}

The early history of computer games was dominated by games being used for educational purposes, or to showcase and demonstrate the new technology. It is natural to consider the same approach for quantum computing. There have been multiple recent gamified approaches for quantum physics including animations ~\cite{Kohnle2010}, games that help to teach quantum concepts~\cite{Cantwell2019,Nita2020,Heider2020} and even game-based citizen-science projects which help participants contribute to research~\cite{Wootton2017}. in this spirit, and making use of the interactive nature of the textbook, a chapter is included for educational games and demonstrations.

\subsection{Interactive demos}

Interactive demonstrations are not limited to this chapter, but can also be found throughout the textbook. The ability to run all these is provided by the `Interactivity Index' section within this chapter.

A key feature of the textbook is that it uses Qiskit to provide practical demonstrations and experiments that cannot be found in traditional textbooks. The benefit of learning through Qiskit over traditional textbooks is that students can easily test their own hypotheses about the behaviour of quantum computers, and shorten the feedback loop between hypothesis, experiment and result.

To enhance this further, the textbook website (through \emph{Thebe}~\cite{thebelab}) enables readers to run Qiskit code directly on the website without needing to install Python, Qiskit, or any other required packages, and with relevant code already run. In a reading group, it was noticed that readers would often like to see the effect of changing a variable, or to further uncover the behaviour of the examples by printing values at different points throughout the code. Without the active code, this would require setting up a separate environment, installing any necessary dependencies, copying the correct code from previous points in the notebook to ensure the environment is setup correctly and all functions and variables are correctly defined, just so the reader can see the effects of a small change.
To allow this behaviour, small changes in the way code is presented had to take place. For example, we avoided splitting logical blocks of code into separate cells, as changing and running multiple cells can lead to confusion and errors; readers would change a line in one cell, but forget to run preceding cells that reset variables required for that logical block. Overwriting behaviour of defined functions was also avoided, so that their definitions do not change at different points in the page. This avoids unexpected behaviour when running cells in different orders.

Finally, interactive python code allows for the integration of widgets through Jupyter widgets~\cite{ipywidgets}. These widgets can be created quickly and make use of Qiskit to simulate quantum computers. One example is the Deutsch-Jozsa algorithm widget. This allows readers to easily simulate the different steps of the Deutsch-Jozsa algorithm, and follow along with the state of the quantum computer at points in the algorithm. The steps can be applied in any order, which encourages readers to recall the steps of the algorithm, but more importantly to apply them in different orders and observe the effects. The number of qubits in the algorithm, and the problem oracle can both be changed to provide many different instances of the same problem.

\subsection{Hello Qiskit}

The game \emph{Hello Qiskit} is a variant of the app the \textit{Hello Quantum}, which was developed as collaboration between IBM and the University of Basel and specifically engineered to provide a gamified introduction to the IBM Quantum Experience.

\textit{Hello Quantum} was intended as a simple puzzle game that can be played by anyone, without any need for prior knowledge about quantum computing. It is based on a visualization of two qubit states and gates, which provide an alternative to linear algebra for reasoning about and understanding two qubits. By completing the game, and by revisiting puzzles to attempt optimal solutions, the player would naturally develop an intuition for how the game is played. This intuition can then be used as a starting point for explaining the quantum context behind the game, which is done by supplying background materials~\cite{hello-quantum} to those players who are interested in learning more. These materials explain how the game states and operations relate to those of qubits, and aim to turn the player's game-based intuition into concrete knowledge of quantum computing.  As such, though the game itself does not teach anything about quantum computing, it creates a context to help people learn. 

For \emph{Hello Qiskit}, the version that is included in the textbook, the same visualization is used and the aim is still to provide the player with useful intuition. However, explanations are interspersed throughout the puzzles rather than being left to the end. Specifically, the puzzles typically have an intro and outro to first explain what will be tackled in the puzzle and then consolidate this after the puzzle is complete.

The topics covered by \emph{Hello Qiskit} reflect many of those covered in the first two chapters. It starts with basic notions about bits, and then introduces their quantum version. It also introduces a universal set of single and two qubit gates. For the case of non-Clifford gates, a modified form of the \textit{Hello Quantum} visualization is used which makes a link to the well-used Bloch sphere visualization. The puzzles provide an opportunity for players to understand common tricks for combining quantum gates, such as converting between \texttt{cz} and \texttt{cx} gates, turning around \texttt{cx} gates and making \texttt{swap} gates. Throughout, the player is able to see the circuit visualization of the circuits they are creating in the game and standard Qiskit terminology is used. The game then concludes with an explanation of Bell's inequalities, which guides the player to create their own experiment to demonstrate the unique nature of quantum variables.

\subsection{QiskitGameEngine}

Another way to learn computation through games is to actually make a game. Creating a simple game can provide an interesting and engaging first task when learning about any new computing concepts~\cite{Renton2016}. For that reason, the textbook includes a very simple game engine, inspired by other simple or education-orientated systems such as the BBC Micro:Bit~\cite{Sentance2017}, PewPew~\cite{pewpew} and PICO-8~\cite{pico8}. The games run directly in the textbook, with Jupyter widgets providing both the screen and buttons. Games are written by addressing individual pixels and buttons. This results in games that are necessarily very limited, with very simple graphics and controls. These limitations therefore encourage the reader to focus on finding clever ways to use Qiskit in the games, rather than being distracted by complex graphics or gameplay.

To introduce this game engine, a section in this chapter gives a simple example of procedural terrain generation using quantum computation, which is an application that has recently begun to be explored~\cite{Wootton2020a,Wootton2020b}. The method presented in this case is one designed specifically for education: it uses just a single qubit, and therefore serves as a hands-on way of using and understanding single qubit gates.

\section{Usage data}


Using data on views and referrals for each page within the textbook, we can build up an idea of how it is being used. Due to an expansion of the textbook that concluded in summer of 2020, we will focus on data from the second half of the year. Within that timeframe is the \emph{Qiskit Global Summer School} in July, as well as the first portion of the academic year in many countries. As much as possible, these will be analysed separately in order to capture different profiles of usage.

\begin{figure}[htbp]
\begin{center}
\includegraphics[width=\columnwidth]{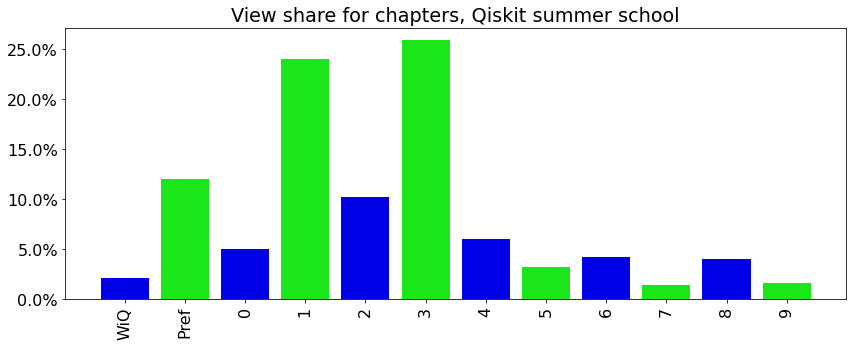}
\includegraphics[width=\columnwidth]{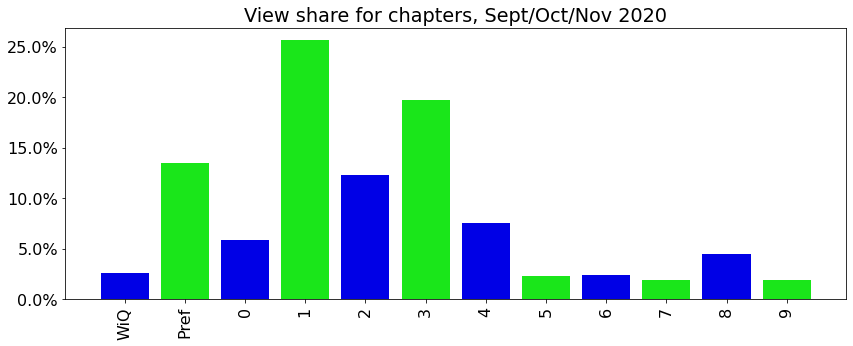}
\caption{These graphs show the percentage of total page views for each chapter. The data for the Qiskit summer school covers 20th-31st July 2020.}
\label{chapter}
\end{center}
\end{figure}

In Fig.~(\ref{chapter}) we see how the views to the textbook are distributed among the chapters, with the `What is Quantum?' section and preface listed separately. The large number of views for the preface is due to this being the landing page of the site. Otherwise, the most viewed chapters are 1 and 3. For the former, this is because Chapter 1 is the beginning of any reader's linear progress throughout the textbook. For the latter, it will be because the algorithms explained in Chapter 3 will be the main point of interest for most readers, presumably leading many to skip straight there or refer back often. The fact that Chapter 1 is the most viewed during the latter part of 2020, whereas Chapter 3 was most viewed during the summer school, is likely due to the summer school offering lectures which paralleled but did not explicitly follow Chapters 1 and 2.

\begin{figure}[htbp]
\begin{center}
\includegraphics[width=\columnwidth]{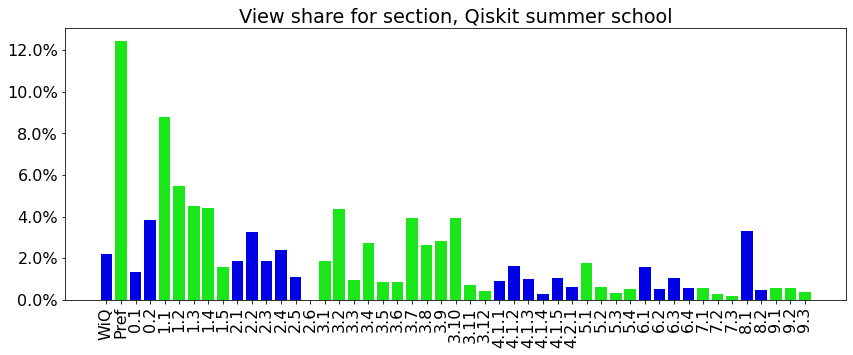}
\includegraphics[width=\columnwidth]{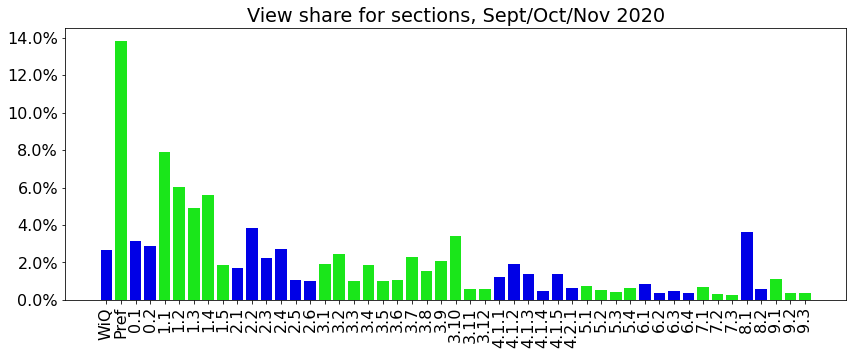}
\caption{These graphs show the percentage of total page views for each section. Chapters are shown in alternating colours for clarity. The data for the Qiskit summer school covers 20th-31st July 2020.}
\label{section}
\end{center}
\end{figure}

These views are further broken down by section in Fig.~(\ref{section}). For the main linear part of the textbook, which runs from Sections 1.3 to 2.5, we see a definite decay of views. This suggests that readers are indeed following this linearly, and so later sections are reached by less viewers than earlier ones. Significant dips are seen at 1.5 and 2.1, likely because these are primarily short motivational recaps before pressing on with the technicalities of Chapter 2.

Chapter 3 does not show the same view decay as the linear part, and these sections typically have at least as many views as the end of the linear part. This suggests that not all readers of Chapter 3 have read the linear part, and that they do not read Chapter 3 linearly. A lack of linear reading is also seen in the later, more technical chapters.

\begin{figure}[htbp]
\begin{center}
\includegraphics[width=\columnwidth]{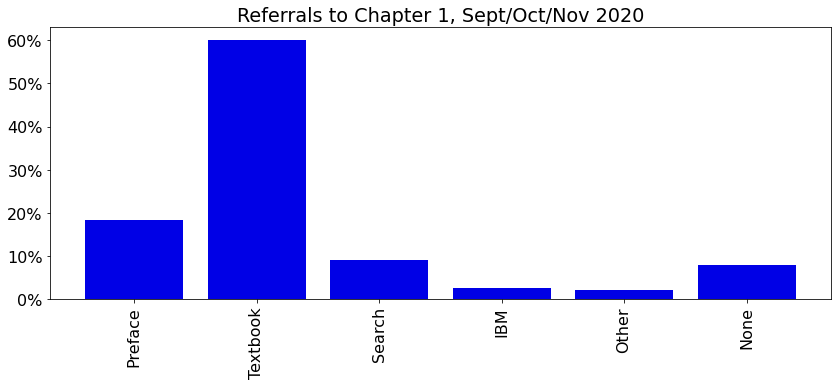}
\includegraphics[width=\columnwidth]{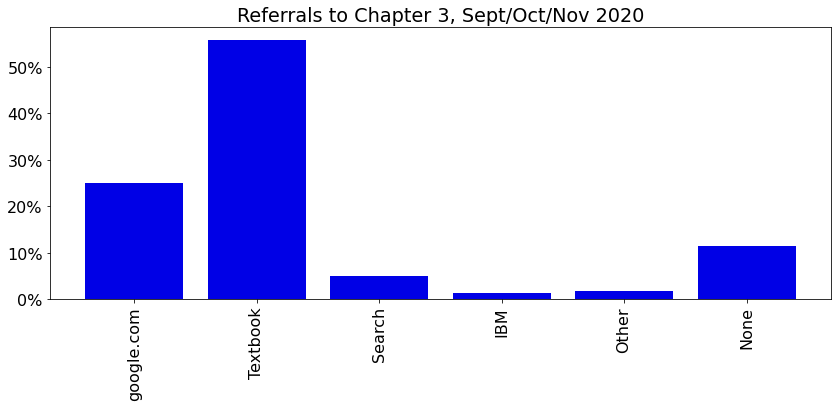}
\includegraphics[width=\columnwidth]{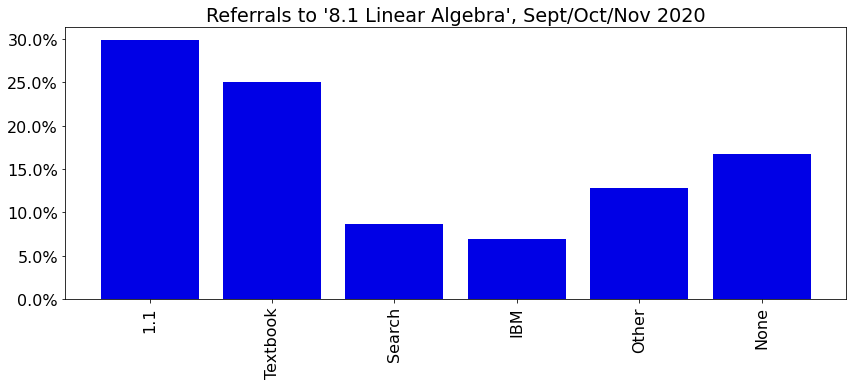}
\caption{These graphs show types of referral to three parts of the textbook (Chapter 1, Chapter 2 and Section 8.1). In each case, the single URL with the most referrals is shown separately to others of the same type.}
\label{referral}
\end{center}
\end{figure}

This can also be seen in Fig.~(\ref{referral}), which shows the referral data for views of certain sections. Though most referrals come from readers navigating the textbook itself, around a quarter of referrals to Chapter 3 come from search engines. Indeed, the single URL that makes the most referrals is google.com. This is in start contrast to Chapter 1 for which around three quarters of referrals come from the textbook itself, with the preface as the single biggest contributor.

Another notable peak in Fig.~(\ref{section}) is that for 8.1, the appendix on linear algebra. Most referrals to this come from the introduction to Chapter 1, which links to the appendix and states ``This chapter will be most effective for readers who are already familiar with vectors and matrices. Those who aren't familiar will likely be fine too, though it might be useful to consult our Introduction to Linear Algebra for Quantum Computing from time to time.'' Some of these views may therefore come from readers who open 8.1 alongside 1.2 when progressing through. This section also has a high number of referrals from IBM-related domains, due to it being a recommended resource for some online events. This section is therefore a significant part of the textbook, as it seems that many either use it as a short cut to to later chapters, or refer back to it for a recap of linear algebra

\begin{figure}[htbp]
\begin{center}
\includegraphics[width=\columnwidth]{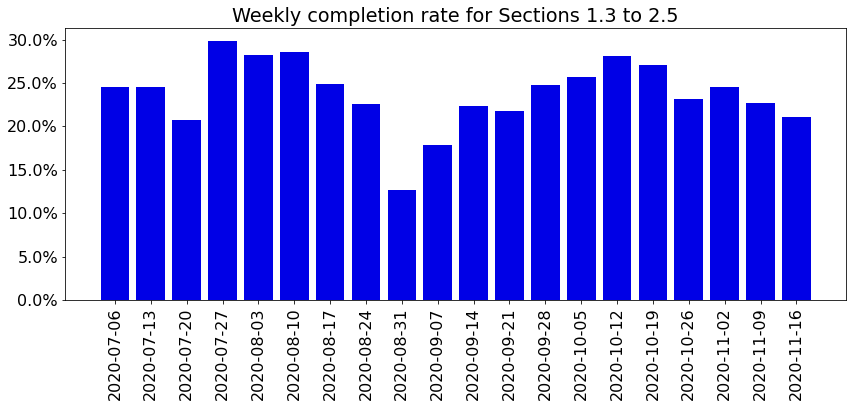}
\includegraphics[width=\columnwidth]{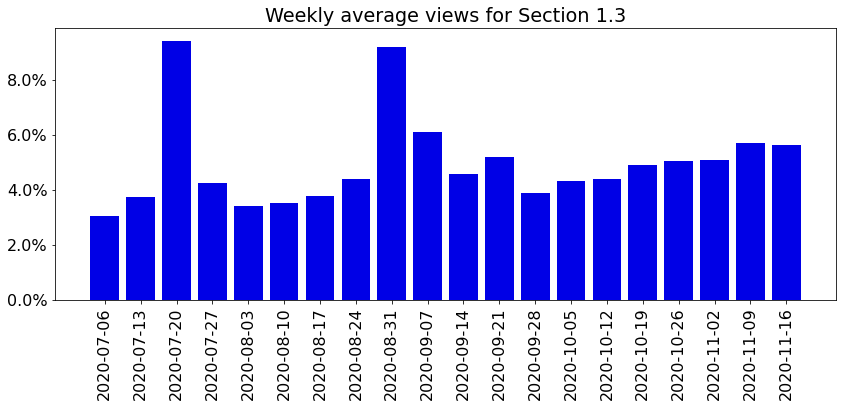}
\caption{The top graph shows the ratio of views for Section 2.5 to views for Section 1.3 for weeks from July-November 2020. The bottom graph shows the total number of views for section 1.3 in those weeks, expressed as a percentage of the total views for the period.}
\label{ratio}
\end{center}
\end{figure}

Since the introductory part from Sections 1.3 to 2.5 is expected to be followed linearly, it would be interesting to determine the completion rate. As an estimate for this, we can look at the ratio of views for its final section to its initial section. This is shown in Fig.~(\ref{ratio}) using weekly averages from July-November 2020. The total number of views for section 1.3 in those weeks is also shown. The ratio is typically around 20\%, but interestingly shows dips during the same weeks as there are peaks for the views of Section 1.3. These are correlated with two major online events which referred to the textbook but did not require this the 1.3-2.5 introduction to be followed. It is not known how this value compares to introductions of similar length in other online textbooks. However, it can be noted that the ratio of 20\% is similar to the read ratio of many of the most viewed articles within the Qiskit blog, such as the background information for \emph{Hello Quantum}~\cite{hello-quantum}, despite the textbook being a much longer and more technical text.

\section{Conclusions}

The textbook `Learn Quantum Computation using Qiskit' has been used by many people since its initial launch in mid-2019. These have included self-learners, lecturers and organizers of events based on quantum computing. Though we have not revealed specific usage numbers here, they are at a level which convinces us to continue and expand the hands-on quantum software-based approach used.

One of the main principles behind the textbook was to not assume complete fluency with the required linear algebra. A tailored introduction to the required concepts is therefore introduced in Sections 1.3 to 2.5. The evidence in the usage data that a significant proportion of readers do indeed go through these sections linearly shows that this approach is indeed well-justified. However, the significant number of views for the linear algebra appendix in Section 8.1 also shows that a significant proportion of readers wish to attempt the short cut. This demonstrates that any attempt to teach the basics of quantum software must explicitly cater both those who are fluent in linear algebra, as well as those who wish to learn as they go.

The usage data also shows that improvements can be made in the dynamics of the textbook: Such as improving the completion rate of the linear introductory sections, helping guide the reader through the rest of the textbook upon completion of the introductory sections; guiding those who enter later chapters from search engines to the introductory sections, and providing a concrete reading order for readers whose background does not require the introductory sections. For readers following a course based on the textbook, this structure will be provided by the course itself. However, more guidance is required to help different profiles of self-learner.

\section{Acknowledgements}

The textbook is the work of over 100 contributors, as is the Qiskit framework on which it is based. The authors would therefore like to thank the entire Qiskit community, without whom the textbook would not be possible.

\bibliography{references}

\end{document}